\documentclass[a4paper,12pt]{iopart}
\usepackage{graphicx}

\def\bea {\begin{eqnarray}}
\def\eea {\end{eqnarray}}
\def\be {\begin{equation}}
\def\ee {\end{equation}}
\def\nn {\nonumber}

\begin{document}

\title{Nuclear modification of charged hadron production at LHC}

\author{Somnath De\footnote[1]{somnathde@vecc.gov.in} and
 Dinesh K. Srivastava\footnote[2]{dinesh@vecc.gov.in}}

\address{Variable Energy Cyclotron Centre, 1/AF, Bidhan Nagar, Kolkata - 700064, India}

\date{\today}

\begin{abstract}
We analyze the recent results  
for suppressed production of charged hadrons
for Pb+Pb collisions at the center of mass energy of 2.76 TeV/nucleon-pair.
We closely follow the treatment used recently by us where partons
lose energy due to radiation of gluons following multiple scatterings
while traversing the quark gluon plasma, before fragmenting into hadrons at 
the center of mass energy of 200 GeV/nucleon-pair. We obtain an empirical value for the
momentum transport coefficient $(\widehat{q})$ and 
provide predictions for azimuthal anisotropy of hadron momenta for non-central collisions.
\end{abstract}

\maketitle


The degradation of energy of high momentum partons or $jets$ inside 
the hot QCD medium manifests itself as a
depletion of particles having large transverse momenta ($p_T$) in
 nucleus-nucleus (AA) collisions at relativistic energies, when compared
 with the corresponding results for proton-proton (pp) collisions.
This phenomenon, often referred to as $\textit{jet quenching}$~\cite{jet_th}, 
is described by the nuclear modification factor $R_{AA}$:
\be 
R^{h}_\textrm{AA}(p_T,b)=\frac{d^2N_\textrm{AA}(b)/dp_Tdy}
{T_\textrm{AA}(b)(d^2\sigma_\textrm{NN}/dp_Tdy)}~,
\ee
where the numerator gives the inclusive yield of hadrons in AA collisions 
for the impact parameter $b$ and the denominator gives the inclusive cross-section
of hadron production in pp collisions scaled with the nuclear overlap function
 $T_\textrm{AA}(b)$.

According to perturbative QCD, the production cross-section of a hadron 
h having a large transverse momentum in pp collisions is written schematically as:
\bea
\frac{d\sigma^{AB\to h}}{dp_Tdy}\sim f^A_i(x_1,\mu_F^2)\otimes f^B_j(x_2,\mu_F^2)\otimes \nn
 \sigma^{ij\to k}(x_1,x_2,\mu_R^2)\otimes D^0_{k\to h}(z,\mu_f^2)~,
\eea
where $f^A_i(x_1,\mu_F^2)$ is the parton distribution function of the $i$-th parton,
 carrying a momentum fraction $x_1$ from the hadron A and
similarly for $f^B_j(x_2,\mu_F^2)$. The parton-parton cross-section 
$\sigma^{ij\to k}(x_1,x_2,\mu_R^2)$, includes all the leading order 
$\mathcal{O}(\alpha_s^2)$
and next-to-leading $\mathcal{O}(\alpha_s^3)$ processes. 
$D^0_{k\to h}(z,\mu_f^2)$ is the vacuum fragmentation probability of
 the parton $k$ into hadron h
at the momentum fraction $z=p_h/p_k$.  

\begin{figure}
\begin{center}
\includegraphics[width=8.5cm, clip=true]{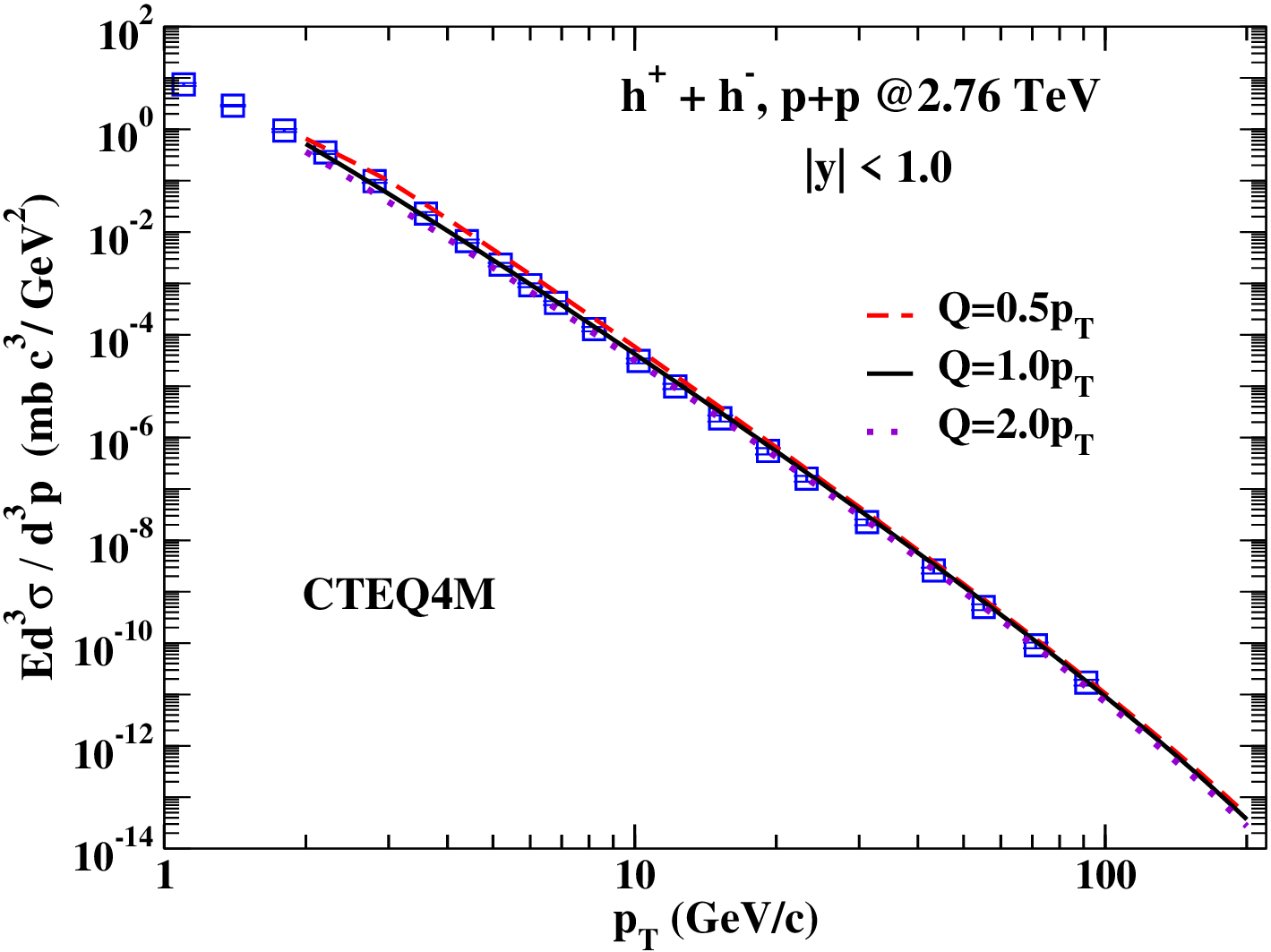}
\end{center}
\caption{(Color online) The differential production cross-section of charged hadrons for p+p
 collisions at 2.76 TeV calculated using
NLO pQCD and compared with the estimates by the CMS Collaboration~\cite{cms_pp}.}
\label{fig1}
\end{figure}

Continuing our earlier study of jet quenching~\cite{sds_jet} at the top Relativistic Heavy
Ion Collider energy (RHIC) ($\sqrt{s_{NN}}$= 200 GeV),
 we analyse the recent results obtained at the  Large Hadron Collider (LHC) at $\sqrt{s_{NN}}$= 2.76 TeV. 
As a first step we show our results for the $p_T$ dependence of   
production cross-section of the charged hadrons for pp collisions at 2.76 TeV 
(Fig~\ref{fig1}).
We have used the same set of structure functions
 ($CTEQ4M$)~\cite{cteq} and fragmentation functions ($BKK$)~\cite{BKK} as before. 
We give results for the
factorization ($\mu_F$), renormalization ($\mu_R$) and fragmentation
($\mu_f$) scales as equal to $Q$, with $Q = 0.5p_T$, $p_T$, and 2.0$p_T$ 
and use NLO pQCD~\cite{aurenche}. The "data" are the estimates used by the CMS Collaboration in 
these studies~\cite{cms_pp}. Buoyed by this success we have used the scale $Q=p_T$ for the
subsequent studies.

Next we have calculated the inclusive production of charged hadrons for Pb+Pb collisions at $\sqrt{s_{NN}}$= 2.76 TeV, accounting
for the multiple scattering and energy loss of partons inside the medium and nuclear shadowing.
 We have used the EKS98 parameterization~\cite{eks98} of nuclear shadowing.

We use the Wang-Huang-Sarcevic model~\cite{wang_frag} of multiple scattering as before and assume that
the probability that a parton traversing a distance $L$ undergoes $n$ multiple scatterings
is given by:
\be
P(n,L)=\frac{(L/\lambda)^n}{n!}\,e^{-L/\lambda}~,
\ee
where $\lambda$ is the mean free path of the parton. We have kept it fixed as
1 fm for both quarks and gluons.
The formalism of parton energy loss is adopted from Baier \textit {et al.}~\cite{baier}
 where the light partons are assumed
to lose energy only through gluon bremsstrahlung.
The formation time of a radiated gluon of energy $\omega$ and transverse
momentum $k_T$ is defined as $t_{\textrm{form}} \approx \omega/k_T^2$. The coherence
length can be defined as $l_{\textrm{coh}} \approx \sqrt{\frac{\omega \lambda}{k_T^2}}$. 
Depending on the formation time (or the coherence length) of the radiated gluon, we consider  three different
 regimes of energy loss (see~\cite{sds_jet,baier} for details). 
When the formation time (or the coherence length) is less than the mean free path ($t_{\textrm{form}} < \lambda$) of the parton,
 we are in the Bethe-Heitler (BH) regime of incoherent radiation. 
The energy loss per unit length in this regime is proportional to the energy
of the parton (E):
\be
-\frac{dE}{dx}\approx\frac{\alpha_s}{\pi}N_c\frac{1}{\lambda}E~,
\label{BH}
\ee
where $N_c$=3. 
If the formation time (or the coherence length) is greater than the mean free path but less than the path 
length $L$ ($\lambda < t_{{\textrm{form}}} < L$), we have a coherent
emission of gluon radiation over $N_\textrm{coh}$(=$({\omega}/ {\lambda k_T^2})^{1/2}$) 
number of scattering centers. 
This is called the LPM regime and the energy loss per unit length becomes:
\be
-\frac{dE}{dx}\approx \frac{\alpha_s}{\pi}
\frac{N_c}{\lambda} \sqrt{\lambda k_T^2 E}~.
\label{LPM}
\ee
\begin{figure*}
\begin{center}
\includegraphics[width=7.5cm, clip=true]{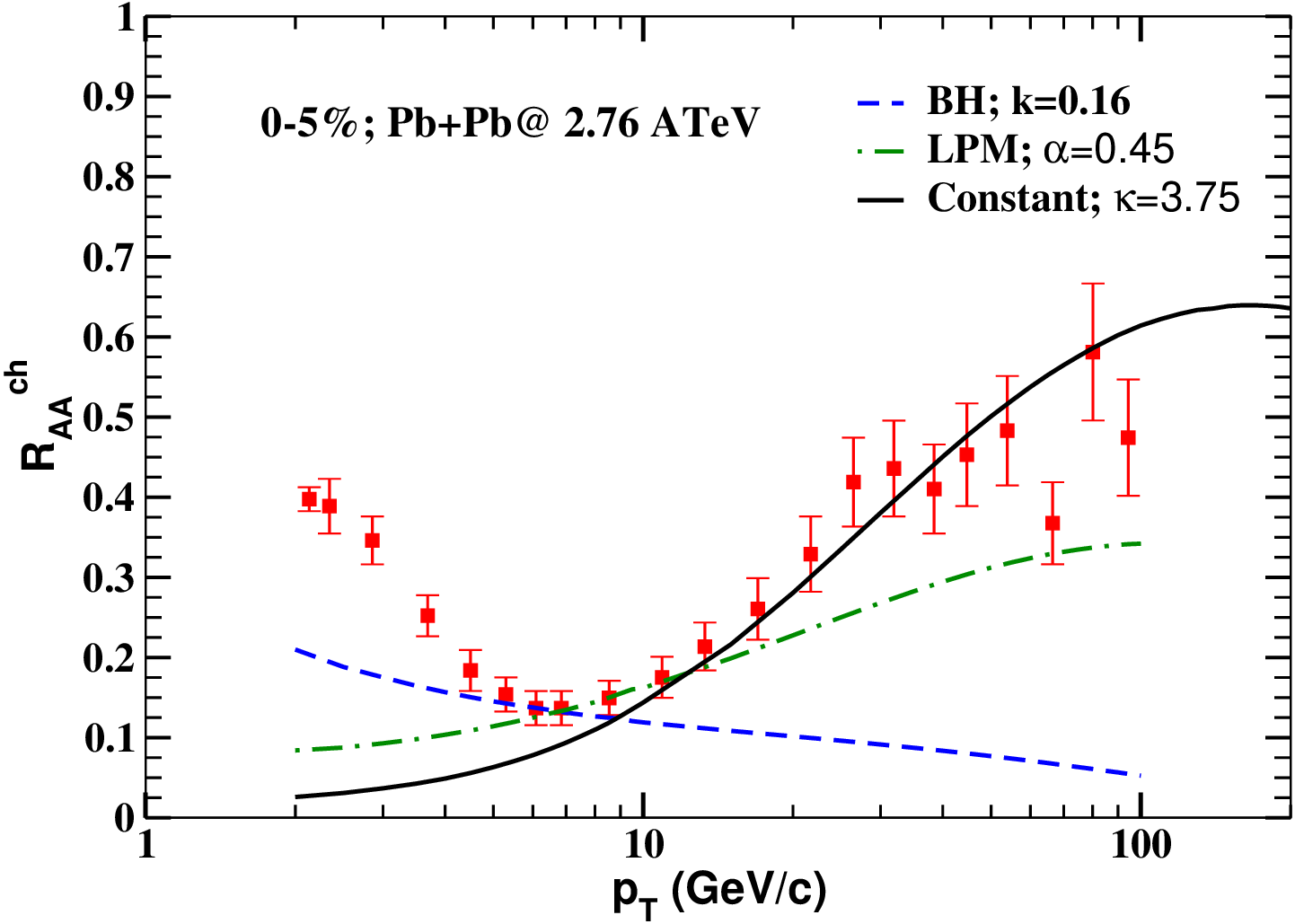}
\includegraphics[width=7.5cm, clip=true]{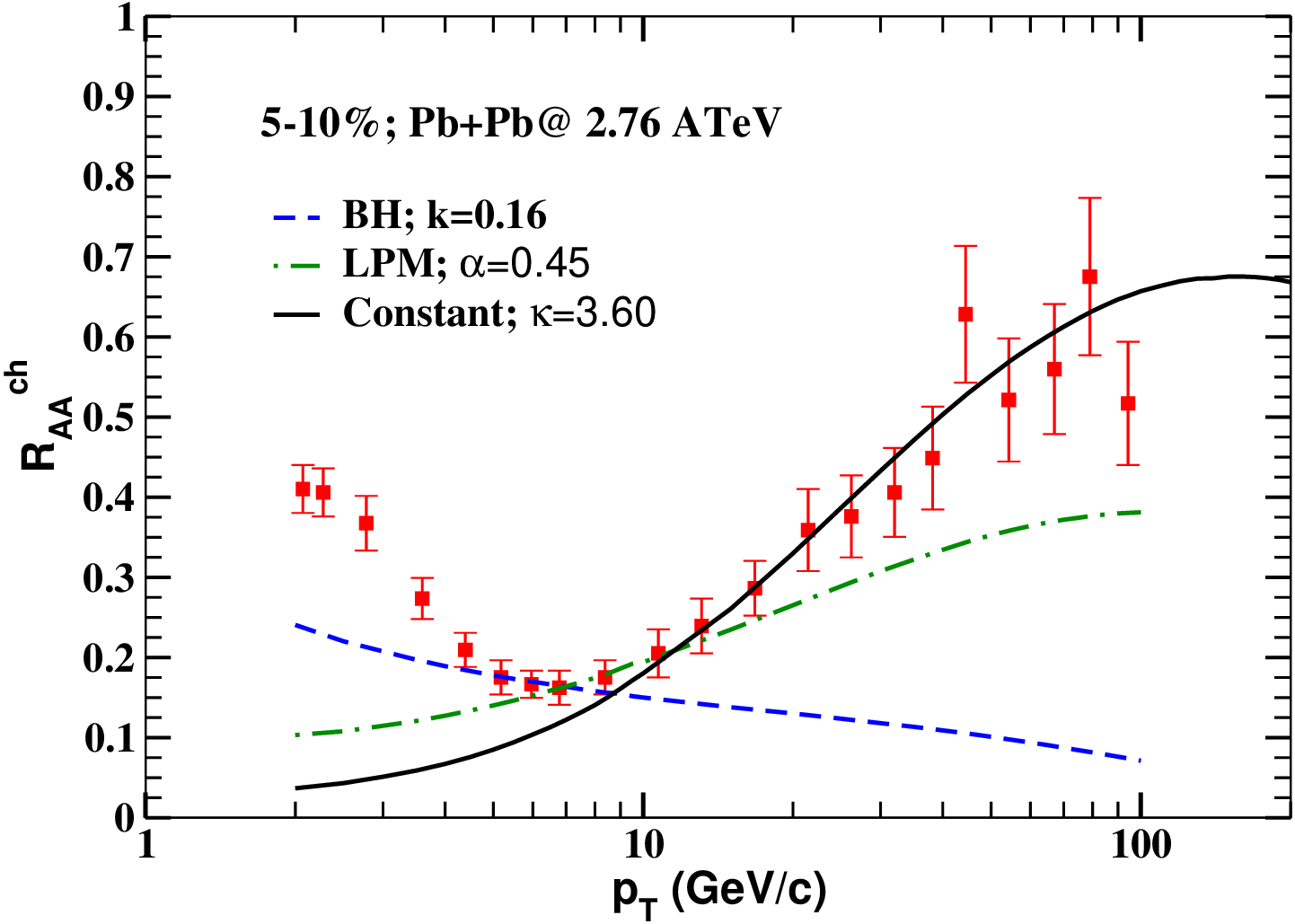}
\includegraphics[width=7.5cm, clip=true]{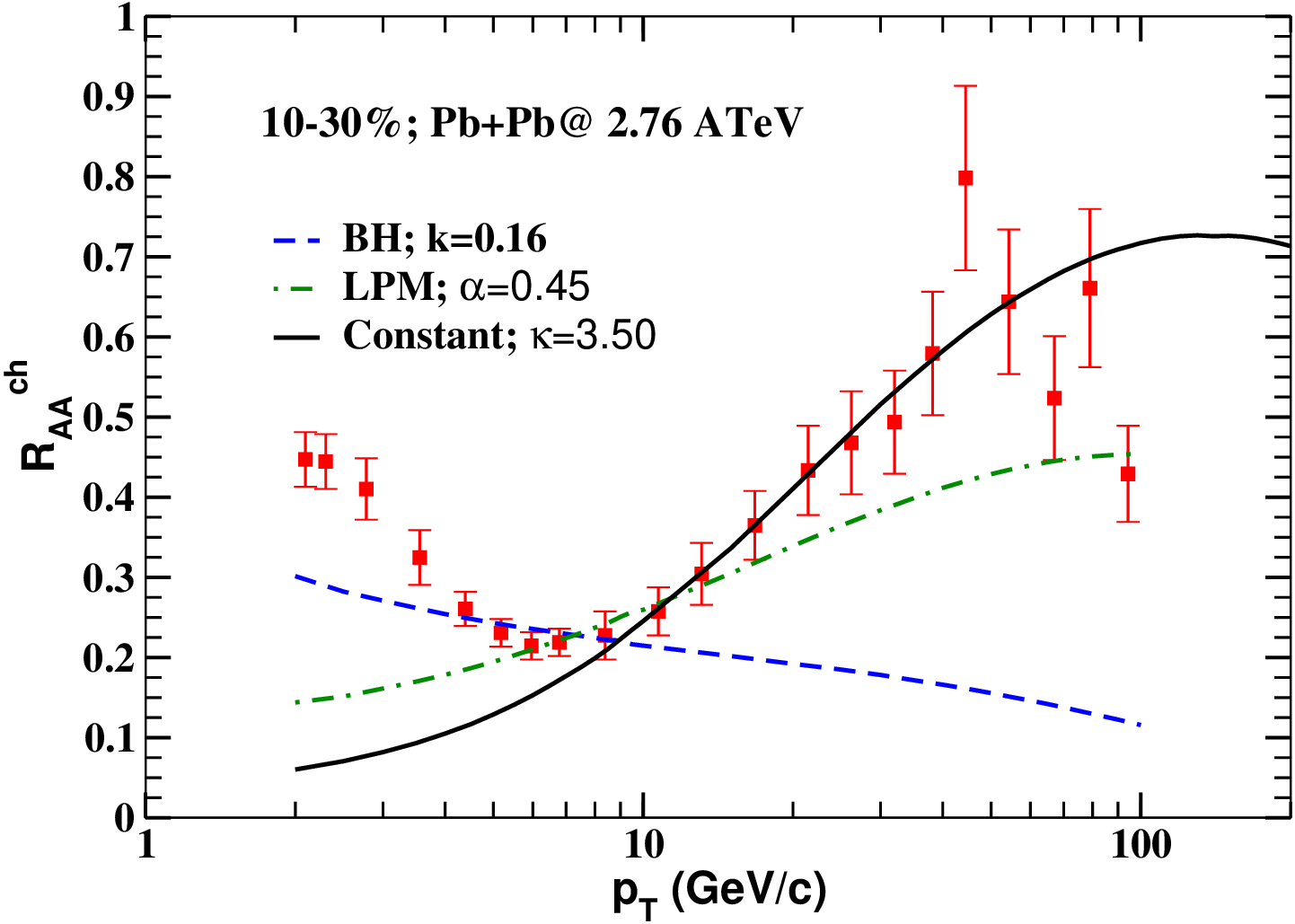}
\includegraphics[width=7.5cm, clip=true]{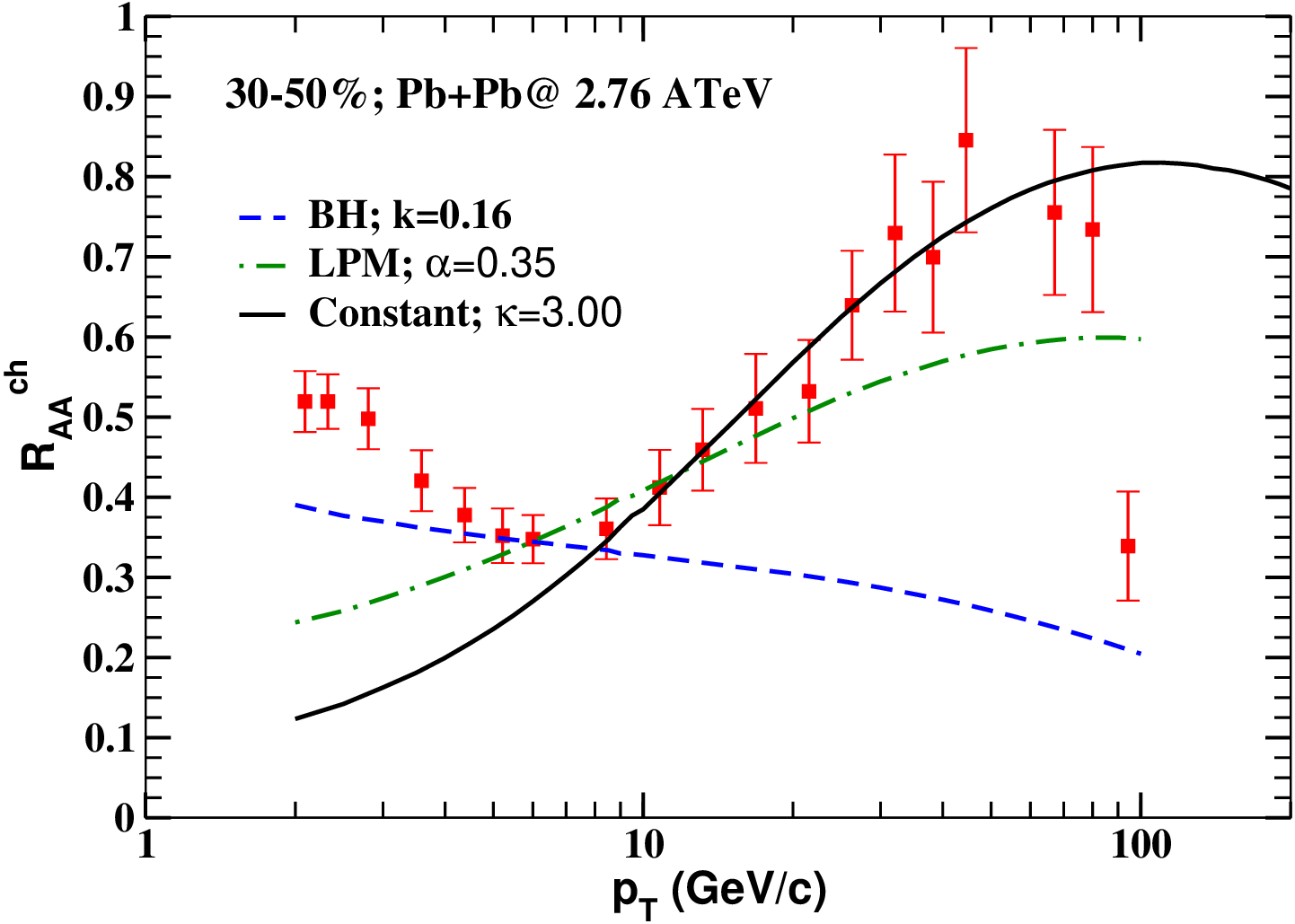}
\end{center}
\caption{ (Color online)
Nuclear modification factor of charged hadron production  calculated for Pb+Pb 
collisions at $\sqrt{s_\textrm{NN}}$= 2.76 TeV, 
in the BH, LPM, and complete coherence regimes of energy loss and 
compared with the measurements by the CMS collaboration~\cite{CMS_jet}.
Note that k is dimensionless, $\alpha$ is in the units of GeV and $\kappa$ has the unit of GeV.}
\label{fig2a}
\end{figure*}

Finally, if the formation time (or the coherence length) is greater than the path length 
$L$ ($t_{\textrm{form}} > L$), we are in the
complete coherence regime of energy loss where the whole medium acts as one coherent source of
radiation. The energy loss per unit length in this regime becomes proportional to the
path length and thus constant for a given value of L.

\be
-\frac{dE}{dx} \approx \frac{\alpha_s}{\pi}N_c \frac{\langle k_T^2 \rangle}{\lambda} L~.
\label{Constant}
\ee
A more careful calculation for Eq.~\ref{Constant} yields (see~\cite{sds_jet,baier}):
\begin{equation}
-\frac{dE}{dx}=\frac{\alpha_s}{4}N_c  \widehat{q} L ~,
\label{qhat}
\end{equation}

where $\widehat{q}$ is the momentum transport coefficient. It may be noted that the coherence (or the absence of it) among the radiated gluons results
in a varying dependence of the energy loss per unit length on the energy (E) of the parton. We shall use this expression to estimate $\widehat{q}$, 
as for these cases, the energy loss per unit length is independent of energy of the parton and it is convenient to compare our results for 
different center of mass energies of the collision and different centralities.
The energy loss per collision, $\varepsilon=\lambda \, dE/dx$, is introduced as a 
free parameter in these studies. We write $\varepsilon$=kE, $\surd (\alpha E )$,
or $\kappa$ for BH, LPM, and complete coherence regimes of energy loss respectively,
where k is dimensionless, $\alpha$ is in the units of GeV and $\kappa$ has the unit of GeV.
The parameters k, $\alpha$, $\kappa$ are varied to get an
accurate description of the nuclear modification factor of charged hadrons 
(R$^\textrm{ch}{}_{\textrm{AA}}$) at different centralities of collisions.
 The average path length $\langle L\rangle$ of the parton inside the medium for 
a given centrality is calculated using the optical Glauber model (see Ref.~\cite{sds_jet}).
 
The energy loss of the parton affects the particle production through the modification of the vacuum
fragmentation function. Following Ref.~\cite{wang_frag}, we write the modified fragmentation function as: 
\bea
 zD_{k\to h}(z,\langle L\rangle,Q^2) =\frac{1}{C_N}\sum_{n=0}^N P(n,\langle L\rangle) \times \nn
 \left \lbrack z_n\,D_{k\to h}^0(z_n,Q^2) + \sum_{m=1}^n z_m\,D_{g\to h}^0(z_m,Q^2) \right \rbrack ~,
\label{mod_frag}
\eea
where $z_n=zE_T/(E_T- \sum_{i=0}^n\varepsilon^i)$, $z_m=zE_T/\varepsilon_m$.
The first term represents the hadronic contribution of a leading parton with a reduced energy 
$(E_T- \sum _{i=0}^n\varepsilon^i)$ and the second term represents the hadronic contribution
of the emitted gluons, each having energy $\varepsilon_m$. 
$C_N=\sum_{n=0}^N P(n,\langle L\rangle)$ and $N$ is
the maximum number of collisions suffered by the parton, equal to $E_T/\varepsilon$.
We add that this treatment explicitly accounts for the fluctuations in the number of collisions 
that the parton may undergo in covering a distance L.

\begin{figure}
\begin{center}
 \includegraphics[width=7.5cm, clip=true]{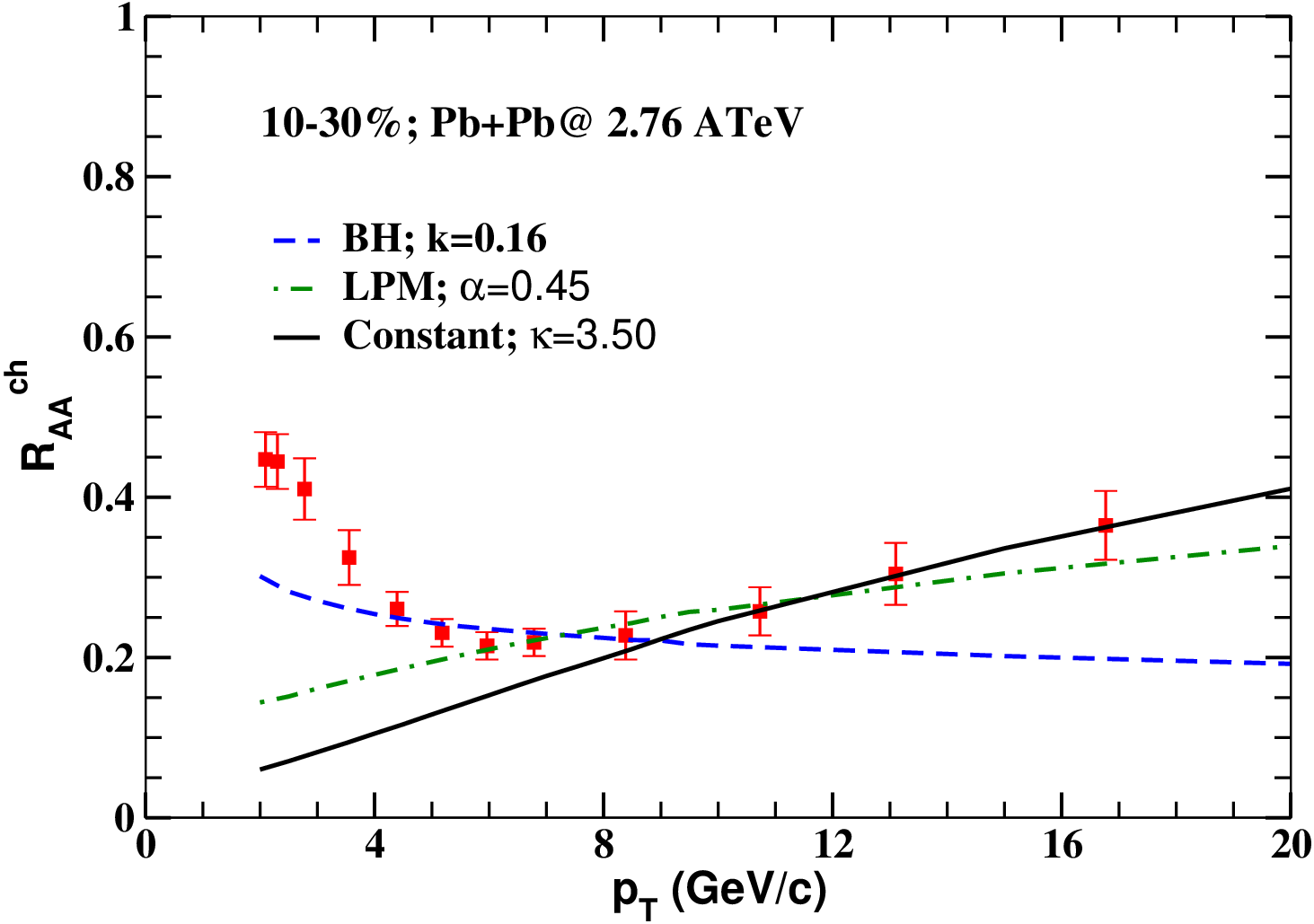}
\end{center}
\caption{(Color online) The charged hadron nuclear modification factor for the centrality class 10-30\% is 
plotted on a linear scale for a smaller $p_T$ range.
Note that k is dimensionless, $\alpha$ is in the units of GeV and $\kappa$ has the unit of GeV.}
\label{fig2b}
\end{figure}
 
The suppressed production of charged hadrons for Pb+Pb collisions at $\sqrt{s_{NN}}$= 2.76 TeV
 is calculated for four 
centralities of collision, namely, 0-5\%, 5-10\%, 10-30\%, 30-50\% for BH, 
LPM and complete coherence regimes of energy loss
(see Ref.~\cite{lhc_jet} for several
other analyses at this center of mass energy).
As mentioned earlier, we have tuned the parameters k, $\alpha$, $\kappa$ systematically 
to have a good agreement with the recent measurement of $R^{ch}_\textrm{AA}$
from CMS collaboration~\cite{CMS_jet}. The results are shown in Fig.~\ref{fig2a}. 

We find the magnitude of suppression can be well described by the BH mechanism 
over a small window of $p_T$, 5--8 GeV/$c$. Of course, at  lower $p_T$ the 
parton recombination dominates over the fragmentation process for particle production at 
RHIC energies~\cite{recomb}.  In addition the hydrodynamic flow 
of the system should affect the hadron spectra for $p_T$ up to 3 GeV/c.
As we go towards higher $p_T$, the BH contribution to nuclear modification gradually drops and
the LPM mechanism is seen to explain the data for the $p_T$ range  $\sim$ (6--15) GeV/$c$
 and even beyond for far-central collisions (30-50\%).

\begin{figure}
\begin{center}
 \includegraphics[width=7.5cm, clip=true]{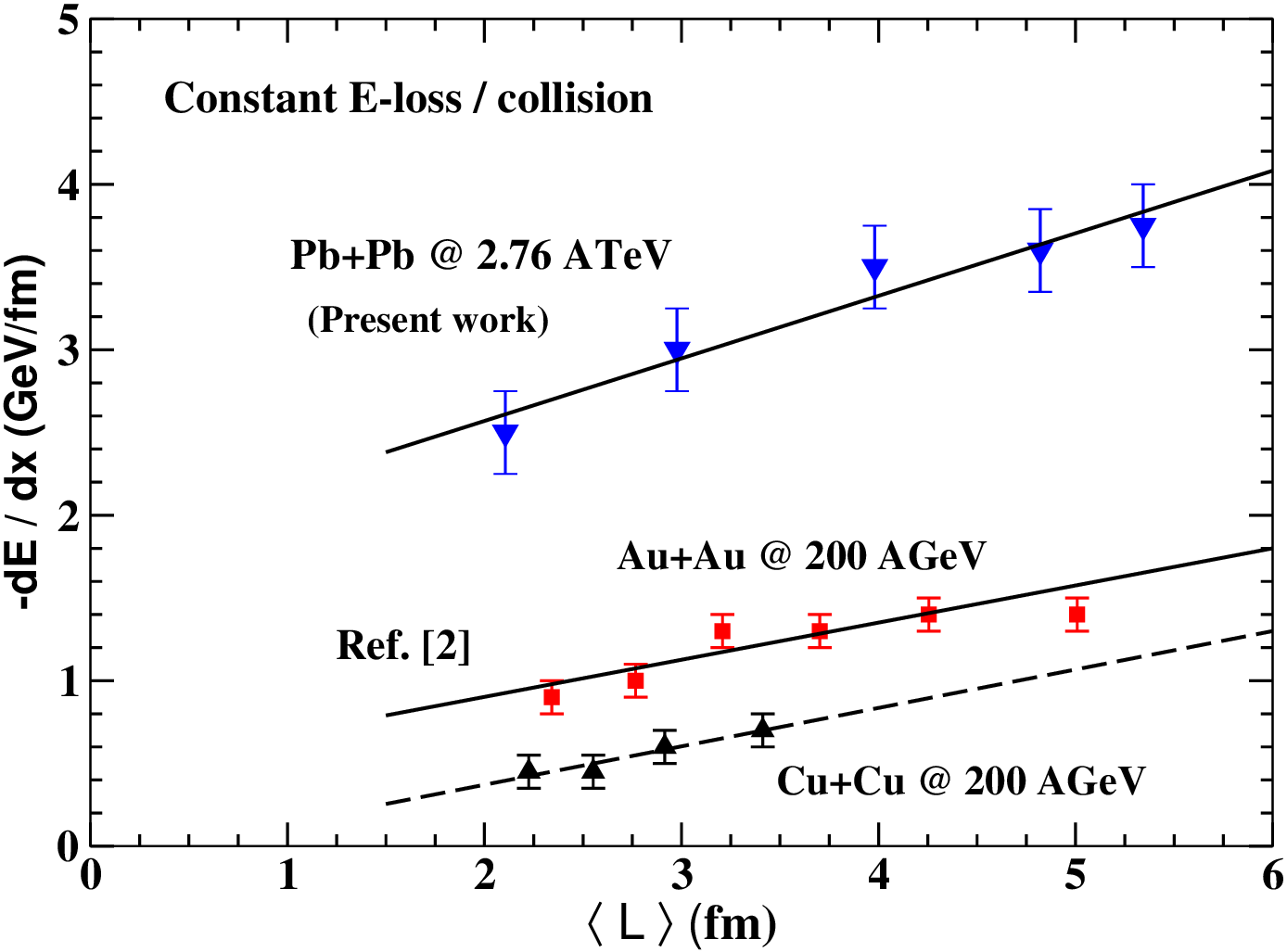}
\end{center}
\caption{(Color online) dE/dx vs average path length, $\langle L \rangle$ of the parton, for Pb+Pb collisions at LHC. 
The results for Au+Au and Cu+Cu collisions are taken from ref~\cite{sds_jet}. 
The corresponding partons have $p_T>$ 8 GeV/$c$ for RHIC energies and $>$ 10--12 GeV/$c$ for LHC energies,
which fall in the complete coherence regime of energy loss.}
\label{fig3}
\end{figure}

 The change in curvature of $R^{ch}_\textrm{AA}$ near 10 GeV/$c$ is correctly followed by 
the complete coherence regime of energy loss.
It gives a good description of data over a broad region of $p_T$; 10 GeV/$c$ $< p_T < $ 100 GeV/$c$.
The curve-over $p_T >$ 100 GeV/$c$ seen in Fig.~\ref{fig2a} seems to have its origin in the antishadowing
in nuclear parton distributions. It will be of interest to repeat this exercise with more recent sets 
of nucleon and nuclear parton distribution functions.
Further, the dominance of the coherent regime of energy loss over the incoherent regime for $p_T>$ (6--8) GeV/$c$
can be seen from Fig.~\ref{fig2b}, where we replot our results in the $p_T$ range (0--20) GeV/c on a linear scale.
Some recent works of jet-quenching~\cite{jeon,dks_jet} have also demonstrated this changing  
mechanism of parton energy loss with $p_T$ for the central collisions.

We note that while the parameter $\kappa$ varies  monotonically with the
centrality of collision (i.e. average path length $\langle L\rangle$),
the parameters $k$ and $\alpha$ are seen to vary only marginally with $\langle L\rangle$. 

It is interesting to study the variation of $dE/dx$ with path length
 for the complete coherence regime, which is seen to work well in the 
range of transverse momenta $\geq$ 10--12 GeV/$c$.
 Recall that we have adjusted the parameter for energy
loss per collision, $\kappa$, for each centrality. 
The linear increase of variation of $dE/dx$ with 
 with the average path length $\langle L \rangle$ for the  Pb-Pb collisions at 2.76 ATeV
is an interesting confirmation of similar findings for the
Au+Au and Cu+Cu collisions at the top RHIC energy in Ref.~\cite{sds_jet} (see Fig.~\ref{fig3}).
Thus the prediction of Baier \textit{et al.} that the total energy loss of the
parton, $\Delta E$ is proportional to $L^2$ is found to be valid at 2.76 ATeV, as well.

We also note that the $dE/dx$ rises more rapidly with $\langle L \rangle$ as the colliding energy increases.
The magnitude of energy loss per unit length of the parton for a given value of $\langle L \rangle$
increases by a factor of 2--3 as we go from RHIC (200 AGeV) to LHC (2.76 ATeV) energy. It remains to be seen
if it rises even more steeply at the top LHC energy at which experimental results are eagerly awaited. 

\begin{figure}
\begin{center}
 \includegraphics[width=7.5cm, clip=true]{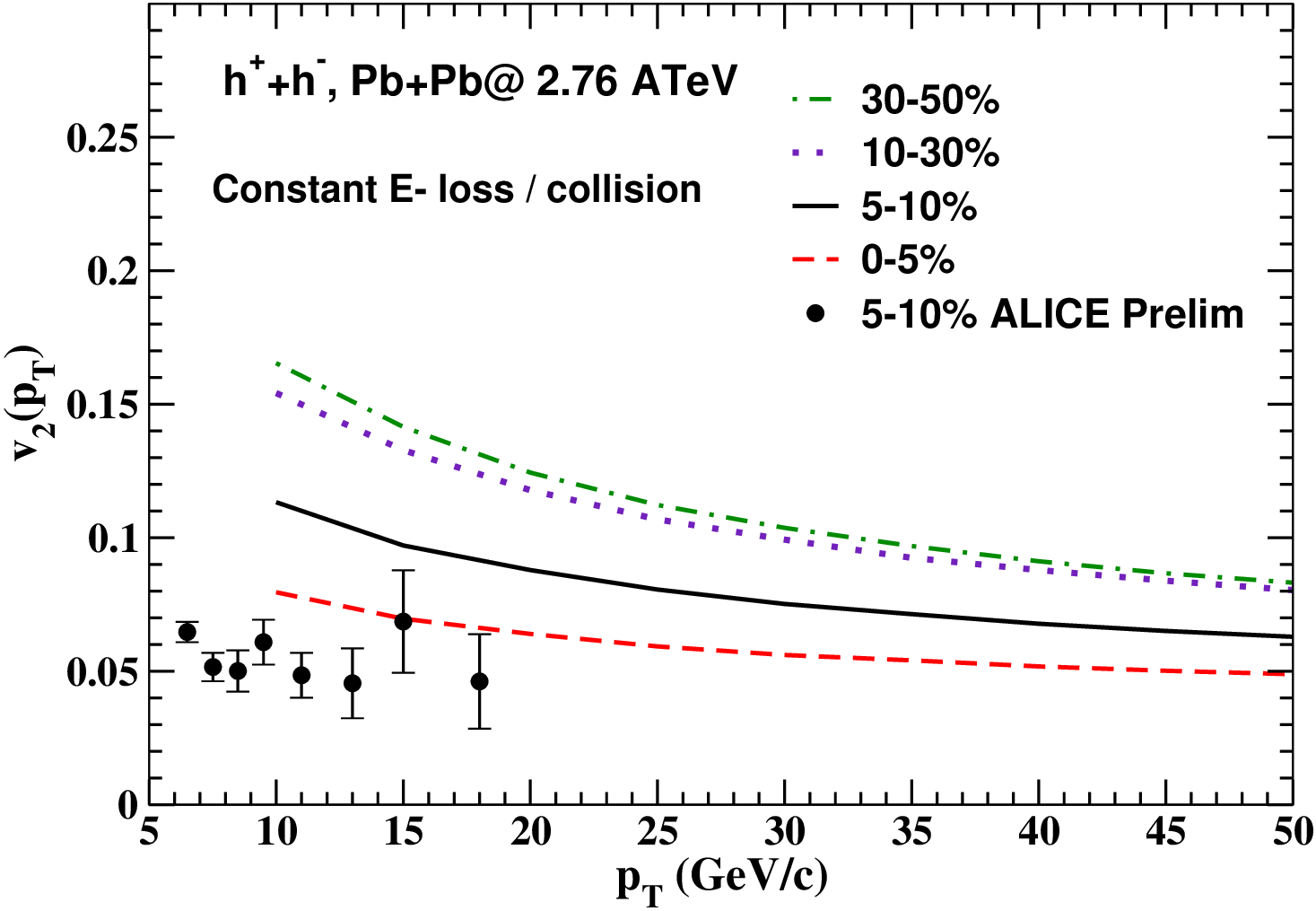}
 \includegraphics[width=7.5cm, clip=true]{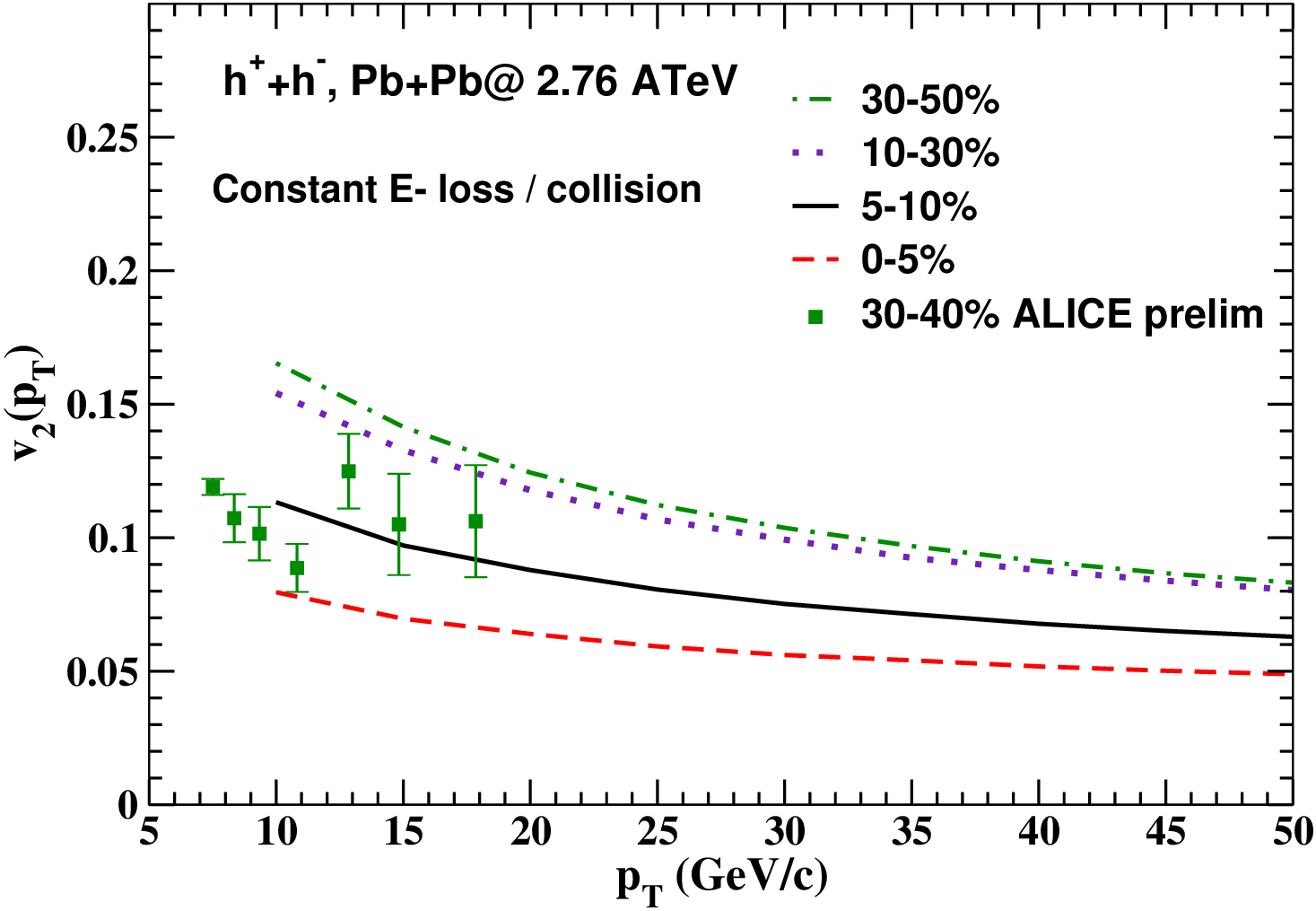}
\end{center}
\caption{(Color online) The differential azimuthal anisotropy coefficient, $v_2 (p_T)$ for the
transverse momentum of hadrons for different centralities, 
using the parameter obtained earlier for Pb+Pb collisions at LHC at 2.76 ATeV. The 
experimental data are from the ALICE Collaboration~\cite{v2_Alice}.}
\label{fig4}
\end{figure}

We can estimate the average momentum transport coefficient $\widehat{q}$  
using Eq.~\ref{qhat} for the QGP medium. 
We find that $\widehat{q}$ varies from 0.63 GeV$^2$/fm for 
0--5\% centrality to 0.91 GeV$^2$/fm for 30--50\% centrality of collisions of Pb nuclei
at 2.76 ATeV. This is about 2.5 times higher than the same obtained for Au-Au collisions at 200 AGeV, 
using a similar analysis~\cite{sds_jet}. 
The smaller value of $\widehat{q}$ at more central collisions may look surprising at first. However
the width of the transverse momentum distribution of the parton, $\langle p_{Tw}^2 \rangle = \widehat{q} \langle L \rangle$~\cite{baier},
can be found to be 3.36 GeV$^2$ for 0-5\% centrality and 2.70 GeV$^2$ for the 30-50\% centrality.

We have also calculated the azimuthal anisotropy of the transverse momentum distribution of
hadrons for $p_T \geq$ 10 GeV/$c$ for non-central collisions as before in Ref.~\cite{sds_jet} and a typical result is shown in
Fig.~\ref{fig4}. We note that our results are larger by a factor of about 2 for the most central
collisions and about 1.5 for less central collisions, when compared to the data from the ALICE Collaboration~\cite{v2_Alice}.
This can perhaps be attributed to uniform density of nuclei and a static medium assumed here.

In brief, we have analyzed the centrality dependence of nuclear modification of hadron
production in collision of lead nuclei at 2.76 ATeV, due to jet quenching. NLO pQCD is used to
generate the distribution of partons which then lose energy by multiple scattering and radiation
of gluons. The formation time of the gluons is used to formulate the effects of coherence, which is 
reflected in different forms of the energy loss per collision. The treatment giving an energy loss
per unit length as proportional to the path length provides a very good description of the data for
$p_T \geq $ 10 GeV/$c$ and leads to a momentum transport coefficient of about 0.6--0.9 GeV$^2$/fm. 
 A comparison with a similar analysis for
collision of gold nuclei at 200 AGeV can be used to suggest a more rapid rise of $dE/dx$ 
with path length and an increased energy loss per collision at the top LHC energy. 

Deviation from this expectation will mean a saturation of energy loss of partons as the temperature of
the medium rises and confirmation will mean that it may or may not appear at higher temperatures.
In either case, the awaited results will be of great interest.


 \end{document}